\documentclass{article}
\usepackage{PRIMEarxiv}
\usepackage{pifont}
\usepackage[utf8]{inputenc} % allow utf-8 input
\usepackage[T1]{fontenc}    % use 8-bit T1 fonts
\usepackage{hyperref}       % hyperlinks
\usepackage{url}            % simple URL typesetting
\usepackage{booktabs}       % professional-quality tables
\usepackage{amsfonts}       % blackboard math symbols
\usepackage{nicefrac}       % compact symbols for 1/2, etc.
\usepackage{microtype}      % microtypography
\usepackage{lipsum}
\usepackage{enumitem} % 编号点包
\usepackage{lastpage}   % 获取总页数
\usepackage{fancyhdr}       % header
\usepackage{graphicx}       % graphics
\graphicspath{{media/}}     % organize your images and other figures under media/ folder
\usepackage{booktabs} % 三线表宏包
\usepackage{longtable}
\usepackage{indentfirst} % 首行缩进宏
\usepackage{indentfirst} % 强制首段缩进
\usepackage{tabularx} % 导言区加载
\graphicspath{{images/}}%自动搜索文件夹图片
\usepackage{titlesec}%斜体
\titleformat{\subsubsection}
  {\normalfont\bfseries\itshape} % 加粗 + 斜体
  {\thesubsubsection}            % 编号格式
  {1em}                          % 编号与标题间距
  {}                             % 标题后内容

\titlespacing{\subsubsection}
  {0pt}     % 左缩进
  {1ex}     % 标题前间距
  {0.5ex}   % 标题后间距
\usepackage{float}%居中函数包
\usepackage{listings}%插入--
\usepackage{xcolor}
\fancypagestyle{plain}{%
  \fancyhf{\headrule} % 清空默认设置
\fancyfoot[LO]{Dongyang Lyu et.al.: \itshape Preprint Submitted to Arxiv}
\fancyfoot[Ro]{Page \thepage\ of \pageref{LastPage}} % 页脚居中显示
}

\title{Construction of Urban Greenland Resources Collaborative Management Platform}

\author{
  Dongyang Lyu \\
  School of Cyberspace Security\\
  Hainan University \\
  Haikou\\
  \texttt{dongyanglv01@gmail.com} \\
   \And
  Xiaoqi Li \\
 School of Cyberspace Security \\
  Hainan University \\
 Haikou\\
  \texttt{csxqli@ieee.org} \\
    \And
    Zongwei Li \\
 School of Cyberspace Security \\
  Hainan University \\
 Haikou\\
  \texttt{lizw1017@gmail.com} \\
}

\begin{document}
\maketitle
\thispagestyle{plain}
\begin{abstract}
Nowadays, environmental protection has become a global consensus. At the same time, with the rapid development of science and technology, urbanisation has become a phenomenon that has become the norm. Therefore, the urban greening management system is an essential component in protecting the urban environment. The system utilises a transparent management process known as" monitoring - early warning - response - optimisation," which enhances the tracking of greening resources, streamlines maintenance scheduling, and encourages employee involvement in planning. Designed with a microservice architecture, the system can improve the utilisation of greening resources by 30\% , increase citizen satisfaction by 20\%, and support carbon neutrality objectives, ultimately making urban governance more intelligent and focused on the community. The Happy City Greening Management System effectively manages gardeners, trees, flowers, and green spaces. It comprises modules for gardener management, purchase and supplier management, tree and flower management, and maintenance planning. Its automation feature allows for real-time updates of greening data, thereby enhancing decision-making. The system is built using Java for the backend and MySQL for data storage, complemented by a user-friendly frontend designed with the Vue framework. Additionally, it leverages features from the Spring Boot framework to enhance maintainability and scalability.
\end{abstract}

% keywords can be removed

\section{Introduction}
Urban growth and rising living standards have made greening essential for improving the quality of life. In China, over 65\% of the population lives in urban areas, yet the average park space per person is only 15.29 square meters,  the recommended 20 square meters \cite{a}. An effective greening management system is crucial for achieving "carbon peaking and neutrality." Access to green spaces can reduce depression risk by 20\% and increase life satisfaction by 5.2 percentage points\cite{b}. Access to green spaces is currently uneven; in cities like Shanghai, central areas have only 60\% of the green space compared to suburban regions. Initiatives such as Shenzhen's "Park City" project aim to address these gaps. Measuring the carbon Benefits of green spaces is vital, with Chengdu reporting a carbon sink value of 1.23 billion yuan. Programs like Beijing’s "Tree Adoption" involve citizens caring for green areas. Innovative greening management systems present practical solutions to urban challenges by enhancing ecological capacity and improving the well-being of residents. This approach contributes to developing more livable cities, fostering a balanced coexistence between urban environments and natural ecosystems\cite{d}.
Foreign urban greening management systems in research and development and operation have accumulated some worthy of domestic urban greening management system research and development and reference results\cite{e, jiaoSurveyEthereumSmart2024}. Various domestic urban greening management systems have been designed and developed utilizing different technologies and architectures. These systems typically offer user management, equipment management, and environmental data monitoring functionalities. However, significant room remains for improvement in system intelligence, analytical capabilities, and overall user experience\cite{f, kumarVulnerabilitiesSmartContracts2024}. Secondly, the domestic scientific research on happy urban greening management systems is still faced with the dilemma of fast technical updates and non-uniform standards, which require continuous improvement and upgrading of technology. Increase the technical research and development and standardization\cite{wangEmpiricalAnalysisSmart2026}. It adopts a frontend and backend separation architecture and realizes enterprise data's comprehensive management and query. They also emphasize scalability and user experience, essential for the happy city green management system\cite{g, zhuSurveySecurityAnalysis2024}. This will help improve user experience and efficiency. It realizes the intelligent management of academic affairs information. The design of system functions and architecture is studied \cite {h}.

\section{System Analysis}
The system is organized into three modules: The administrator, gardener, and supplier modules \cite {m, arceriSoundConstructionEVM2024}. The administrator module is designed to configure user settings, manage equipment, and input information regarding the urban greening environment and plant data. It also oversees the management of purchase applications and conducts statistical analyses pertinent to urban greening efforts\cite{caiEnablingCompleteAtomicity2024}. The gardener module allows users to set personal information, manage plant details, and perform maintenance operations. It also includes the functionality to review environmental data to develop appropriate maintenance strategies, as well as to carry out facility maintenance tasks. The supplier module is responsible for providing flowers and trees. Through effective supply management, this module ensures that resources are allocated efficiently to meet the requirements of customers \cite {n}.

\subsection{Feasibility Analysis}
\begin{enumerate}[label=\textbullet]

\item The Urban Greening Management System utilises Spring Boot for the backend and Vue.js for the front end, resulting in an effective and easily maintainable framework. This strategic combination facilitates optimised resource utilisation, leading to a reduction in maintenance costs by more than 30\%. It also extends the lifespan of greening facilities while supporting carbon trading initiatives   \cite {q}. 

\item The system also contributes to a decrease in management and time costs, resulting in an overall expense reduction of 20\% compared to other cities. It effectively meets the ecological needs of residents and encourages public participation through technological advancements. While challenges like the digital divide are present, strategic planning and implementation can effectively address these issues \cite{o}. 

\item Overall, the system enhances ecological quality, fosters social connections, and modernises governance, demonstrating its effectiveness across multiple domains. Therefore, the system is feasible at the economic, technical, and social levels\cite {p}.

\end{enumerate}

\subsection{Requirement Analysis}
\subsubsection{Administrator Use Case}
\begin{figure}[htpb]
  \centering
  \includegraphics[width=0.6\textwidth]{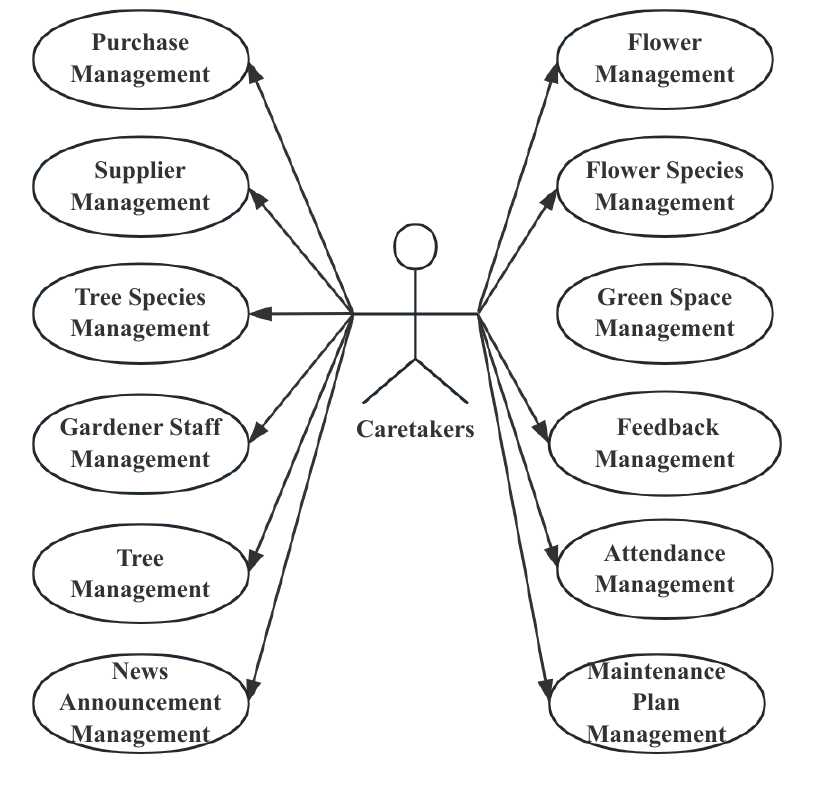}
  \caption{Administrator Use Case Diagram}  % 标题必须在前
  \label{fig:1}
\end{figure}

The administrator oversees the platform's operations, including managing trees, flowers, green spaces, and maintenance programs. The system streamlines the recording, searching, and updating of information, ensuring the effective implementation of urban greening initiatives
(See Figure\ref{fig:1}). A feedback mechanism is integrated to enable administrators to monitor maintenance tasks promptly and address any issues that may arise. This management system enhances operational efficiency and plays a vital role in maintaining the appeal and Health of green spaces.

Administrators should be able to add, delete, and modify relevant details, including names, contact information, and skill levels, to manage gardener information. Furthermore, the system must facilitate the querying of gardeners' work records to ensure the efficient management of the gardening team and enhance the quality of urban greening maintenance.
\begin{itemize}
\item[(1).]\textbf{Tree Management}: Administrators must be able to record critical information regarding trees, including their location, species, growth status, and maintenance history. The system should facilitate queries based on specific conditions to enable the formulation of targeted maintenance plans, ensuring the healthy growth of trees.
 \item[(2).]\textbf{Tree Species Management}: Administrators must manage comprehensive information on tree 1
species, covering names, characteristics, and suitable environmental conditions. The system should support species addition, editing, and deletion, thereby providing a scientific foundation for effective tree planting and maintenance.
\item[(3).]\textbf{Flower Management}: Administrators should monitor each flower's growth cycle, bloom time, and maintenance needs. The system must allow for efficient checking of flower status and facilitate timely adjustments to maintenance strategies to ensure flowers bloom beautifully.
\item[(4).]\textbf{Flower Species Management}: Administrators are tasked with maintaining detailed records of flower species, including names, colors, flowering periods, and other pertinent characteristics. The system must support adding, deleting, modifying, and retrieving species information to provide valuable flower planting and planning data.
\item[(5).]\textbf{Green Space Management}: Administrators need to plan and manage the layout of green spaces, recording their health status and dimensions. The system should analyze the efficiency of green space utilization, optimize design, and enhance the overall aesthetic appeal of urban greening efforts.
\item[(6).]\textbf{Maintenance Plan Management}: Administrators should be equipped to develop and adjust maintenance plans comprising watering, fertilizing, and pruning tasks. The system must track the implementation of these plans to ensure that maintenance activities are completed on schedule and to a high standard.
\item[(7).]\textbf{Implementation Feedback Management}: Administrators must gather and analyze conservation efforts, including task completion and any recorded issues. The system should enable prompt responses to this feedback, allowing for adjustments to conservation strategies and improving overall management efficiency.
\item[(8).]\textbf{Purchase Management}: Administrators are responsible for compiling a list of required resources, which are communicated to suppliers through a cloud-based platform. Supplier feedback is subsequently reviewed by administrators to enhance supply chain efficiency.
\item[(9).]\textbf{Attendance Management}: Administrators should evaluate employee attendance by monitoring sign-in records, thereby improving the efficiency of employee management.
\item[(10).]\textbf{News Announcement Management}: Administrators utilize the backend to disseminate important announcements to staff, with the capability to reiterate reminders for significant events.
\item[(11).]\textbf{Supplier Management}: Administrators should have the ability to select suppliers based on specific criteria, facilitating cost optimization and improvement in the quality of supplies received.
\end{itemize}

\subsubsection{Supplier Use Case}

\begin{figure}[ht]
  \centering
  \includegraphics[width=0.5\textwidth]{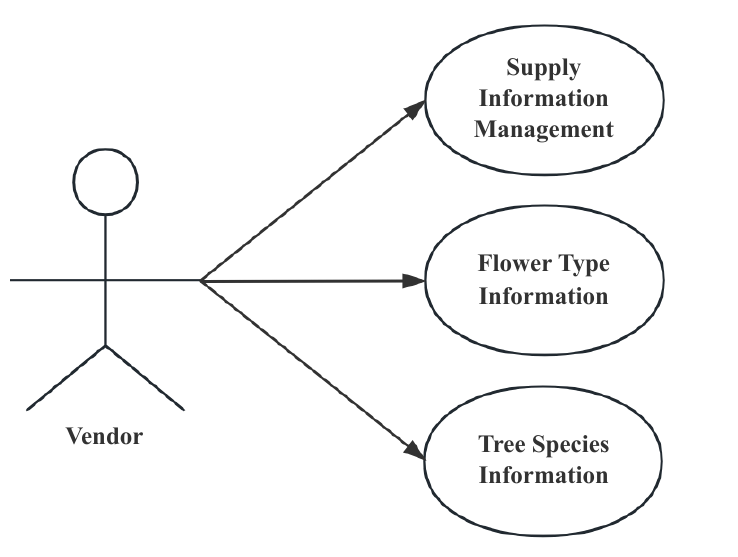}
  \caption{Supplier Use Case Diagram}  % 标题必须在前
   \label{fig:2}
\end{figure}

The supplier is responsible for providing trees and flowers on the platform and managing their associated information. They upload resources to the cloud, enabling users to download and review them conveniently(See Figure\ref{fig:2}). 
The system facilitates suppliers' submission of product data, ensuring that the supply database remains current. Users can search for trees and flowers by name or serial number, allowing efficient access to the desired information. 
\begin{itemize}
\item[(1).] \textbf{Supply Information Management}: Suppliers submit supply information, including product name, quantity, price, and supply time, through the system. The system subsequently validates this information and updates the inventory database accordingly.
\item[(2).]\textbf{Tree Species Information}: Users can retrieve the tree catalogue by entering the name or serial number. The system will then return the corresponding species' growth characteristics and supply requirements.
\item[(3).]\textbf{Flower Type Information}: Users can filter flower varieties based on name or serial number criteria, allowing them to view different categories' current availability and specifications. This feature ensures convenient and timely access to inventory information.
\end{itemize}

\subsubsection{Gardener Use Case}
\begin{figure}[H]
  \centering
  \includegraphics[width=0.7\textwidth]{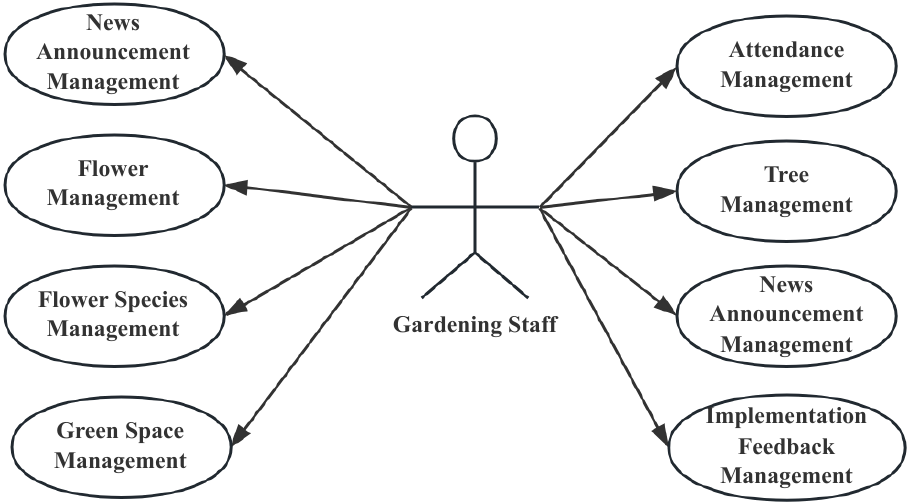}
  \caption{Use Case Diagram For Gardeners}  % 标题必须在前
   \label{fig:3}
\end{figure}

Gardening staff play a critical role in conservation efforts. They manage the maintenance program through the system(See Figure \ref{fig:3}), enabling them to view and execute assigned tasks efficiently and ensuring that all maintenance activities are conducted according to plan. Upon completing their assignments, gardeners can submit detailed feedback, which is instrumental in informing subsequent maintenance efforts. Additionally, the system supports the effective management of trees, flowers, and green spaces, allowing gardeners to document growth status, maintenance history, and other relevant information at any time. This capability significantly enhances the efficiency and quality of maintenance work, ensuring that urban greening initiatives remain vibrant and visually appealing \cite {r}. 

\begin{itemize}
\item[(1).]\textbf{Maintenance Plan Management}: Gardeners should be able to access their assigned maintenance plans, which include tasks such as watering, fertilizing, and pruning. The system must detail the plan by providing information on timing, location, and required tools. Gardeners should also be able to update the task completion status to ensure the timely execution of the maintenance plan.

\item[(2).]\textbf{Implementation Feedback Management}: After completing conservation tasks, gardeners should be able to submit implementation feedback. This feedback should encompass task completion status, any issues encountered, and suggestions for improvement. The system must record this feedback so that administrators can review and adjust the maintenance plan as necessary.

\item[(3).]\textbf{Tree Management}: Gardeners should be able to document trees' growth status, maintenance history, and abnormalities. The system should provide three information query functions to enable gardeners to monitor tree conditions at any time, facilitating targeted maintenance efforts.

\item[(4).]\textbf{Flower Management}: Gardeners are responsible for the daily maintenance of flowers, including watering, fertilizing, and weeding. The system should support entering and retrieving flower information, helping gardeners understand the growth cycles and optimize their maintenance activities.

\item[(5).]\textbf{Green Space Management}: Gardeners must ensure green spaces' tidiness and aesthetic appeal, including litter removal and lawn mowing. The system should provide a green space management interface that displays the area's layout, records changes in health status, and tracks maintenance tasks to assist gardeners in their maintenance efforts.

\item[(6).]\textbf{Flower Species Management}: The system should allow for the recording of various flower types and their associated images. Additionally, it should enable the characterization of flowers so that users can efficiently query specific flower types.

\item[(7).]\textbf{Attendance Management}: Employees should be able to record their attendance through a sign-in feature and access their historical attendance records.

\item[(9).]\textbf{News Announcement Management}: The system should remind employees of essential notes, requirements, and standardized guidelines on the main interface, thereby enhancing the accuracy and quality of their work.
\end{itemize}

\section{System Architecture}
\begin{figure}[htpb]
  \centering
  \includegraphics[width=0.8\textwidth]{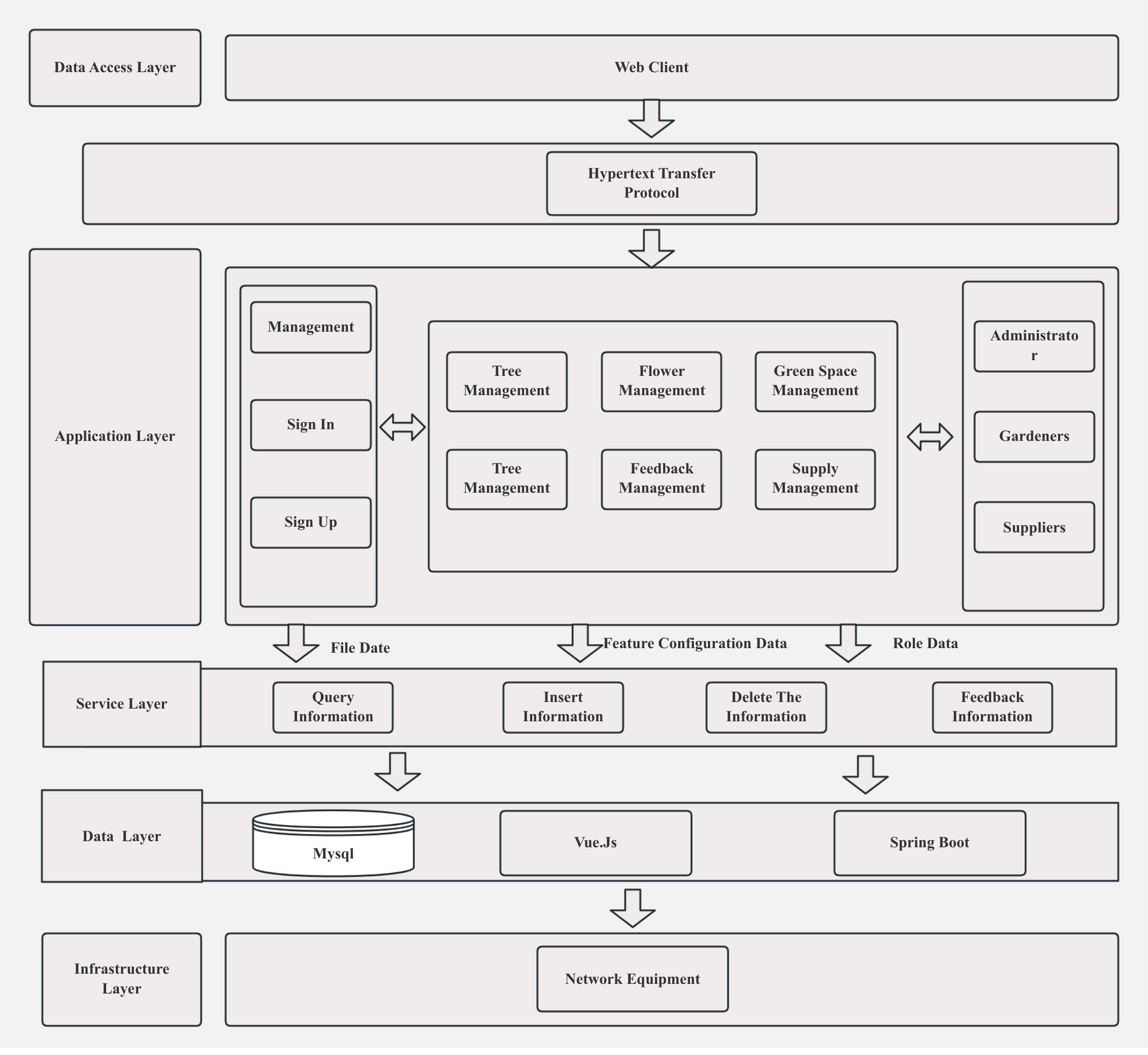}
  \caption{System Architecture Diagram}  % 标题必须在前
  \label{A:2}
\end{figure}

This system uses a layered architecture design to build a modular, highly secure, scalable garden management platform\cite{xiao2025parallelizing}. The architecture is divided into five layers: The Web client layer is based on Vue.js to achieve a responsive front-end, adapting to the 4K resolution display requirements and providing differentiated interactive interfaces for multiple roles, such as administrators, gardeners, and vendors; the application layer relies on the Spring Boot framework to provide core business functions, such as user authentication, vegetation management (trees/flowers/green space), supply chain collaboration, and feedback processing, via RESTful API; and efficient front-end and back-end communication based on HTTP/HTTPS protocols\cite{aguilarSmartContractFamilies2024}; the architecture is divided into five layers(See Figure\ref{A:2}) .The application layer is developed based on the Spring Boot framework. RESTful APIs provide core business functions such as user authentication, vegetation management (trees/flowers/green areas), supply chain collaboration, and feedback processing\cite{i, wang2024ContractsentryStaticAnalysis}. It achieves efficient front-end and back-end communication based on HTTP/HTTPS protocols. The infrastructure layer has high-performance network equipment to support 4K content transmission and real-time business processing in high-concurrency scenarios\cite{sun2025FIRESmartContract}.

Functional design: Administrators can manage the entire supply chain, while gardeners are limited to vegetation maintenance operations\cite{hanOSwapPreservingAtomicity2026}. The core business module encompasses the complete life cycle management of vegetation, including classification entry, status tracking, and data analysis\cite{heCodeNotNatural2024}. It also facilitates closed-loop processing of user feedback, such as problem reporting and work order flow, while optimising the supplier collaboration process. Additionally, it integrates operation logging and security auditing functions to ensure business traceability\cite{huangAdvancingWeb302024}. In terms of non-functional requirements, the system strengthens security through HTTPS encrypted transmission, hash storage of sensitive information, and regular vulnerability scanning; it adopts Redis caching, load balancing, and microservices architecture to improve performance, supporting 1,000+ concurrent requests per second, with a response time of 500ms or less\cite{liASTRODetectingAccess2025}; it achieves elasticity through the deployment of Docker containers and Kubernetes clusters, and ensures horizontal expansion\cite{j, suDiSCoDecompilingEVM2025}. Docker containerisation and Kubernetes cluster deployment enable flexible expansion and reduction of capacity, guaranteeing horizontal scalability; in addition, daily incremental backup of the database and a 30-minute fault recovery mechanism further enhance the system’s disaster recovery capability\cite{y, wangContractCheckCheckingEthereum2024}..

The HTTP protocol serves as the core communication hub in this system. As a standardized interaction protocol between the Web client (Vue.js front-end) and the application layer (Spring Boot back-end), it defines the request-response model\cite{wangEfficientlyDetectingReentrancy2024}. It ensures decoupled collaboration between the front end and back end. It supports the realization of all business functions, including user authentication (login/registration), resource management (trees, flowers, and green spaces), feedback, and supply chain operations, all of which transmit commands and data via HTTP\cite{zhangEVMShieldInContractState2024}. In terms of technical implementation, the front-end initiates HTTP requests (e.g., GET/POST) to the back-end service layer, which triggers the data layer (MySQL) to add, delete, change, and check operations and returns the results in JSON format. At the same time, HTTP integrates permission control (administrator, gardener, and vendor roles) and secure access through the cookie/Token mechanism. As a cornerstone of modern Web applications, the high compatibility and scalability of this protocol provide a stable and efficient communication foundation for the system\cite{boi2024VulnHuntGPTSmartContract}.

Regarding technical implementation, the system follows the principle of front-end and back-end separation, with Vue.js as the core of the front-end and Spring Boot framework in the back-end, and strictly aligns with OAuth2.0 and JWT standards to achieve authentication\cite{hu2023LargeLanguageModelPowered}. The deployment level relies on cloud-native technology to ensure cross-platform compatibility and ease of operation and maintenance\cite{l, wei2025AdvancedSmartContract}. In the end, through the output of complete architectural design documents, interface specifications, and operation and maintenance manuals, the system can efficiently support multi-role collaboration, complex business scenarios, and future functionality expansion needs with stability and flexibility.

\subsection{Database Analysis And Design}
\subsubsection{Detailed Design of Database}
Urban Greening Management System mainly contains seven core components: Administrator module, gardener module, supplier module, flower management, tree management, green space management, and purchase application module. Among them, the administrator module includes the administrator name, user name, password, and other information; the gardener personnel module consists of the gardener name, phone, password, and other information; the supplier module consists of the supplier company name, phone, password, and other information; flower management module includes flower name, photo description and other information; tree management module includes tree name, photo description and other information; purchase application module includes order number, Order details, shipping status and other information \cite{u, ayubSoundAnalysisMigration2024}.

\begin{figure}[H] % [H] 表示强制当前位置
  \centering
  \includegraphics[width=0.8\textwidth]{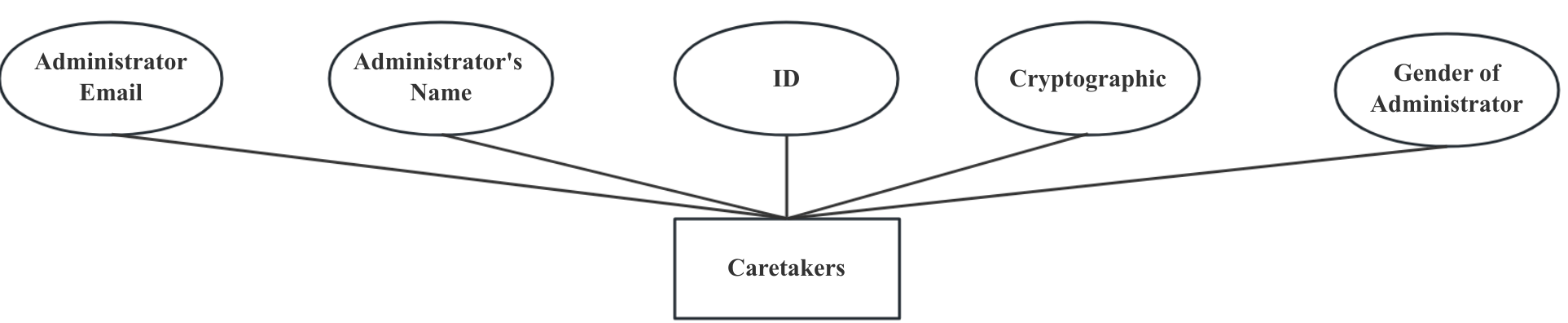}%width可以修改宽度
  \caption{Administrator Entity Property Graph}
   \label{fig:5}
\end{figure}

 (1) \textbf{Administrator Entity}: The administrator entity usually covers the administrator name, password, username, gender, and email attributes(See Figure\ref{fig:5}). 

\begin{figure}[H] % [H] 表示强制当前位置
  \centering
  \includegraphics[width=0.8\textwidth]{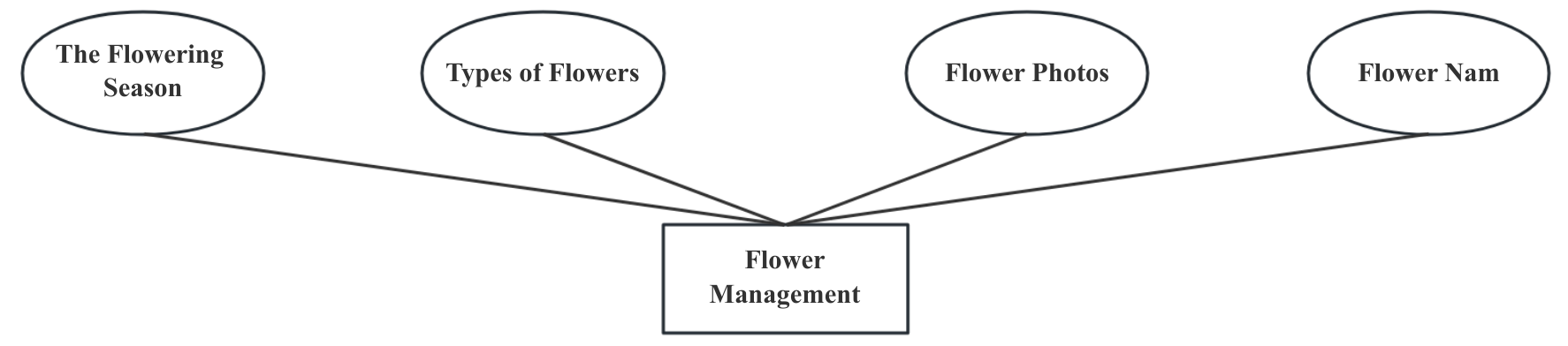}%width可以修改宽度
  \caption{Flower Management Entity Attribute Map}
   \label{fig:6}
\end{figure}

(2) \textbf{Flower Management Entities}: The flower entity is a data object that stores information about the flower. It contains attributes such as flower type, bloom period, name, and picture.(See Figure\ref{fig:6}) 

\begin{figure}[H] % [H] 表示强制当前位置
  \centering
  \includegraphics[width=0.6\textwidth]{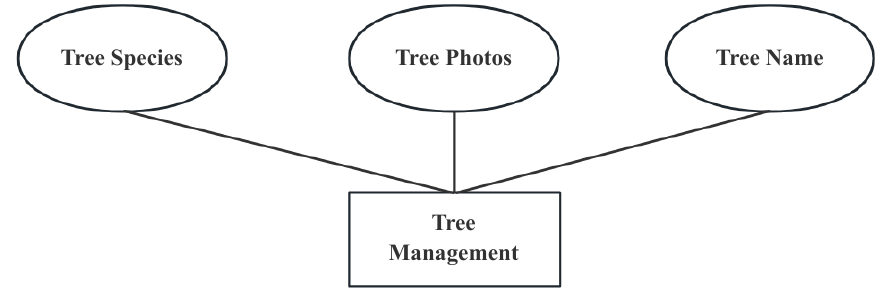}%width可以修改宽度
  \caption{Tree Management Entity Attributes}
   \label{fig:7}
\end{figure}

(3) \textbf{Tree Management Entities}: The tree management entity usually stores the tree type, name, and photo attributes. The following is a diagram of the attributes of the tree management entity(See Figure\ref{fig:7}).

\begin{figure}[H] % [H] 表示强制当前位置
  \centering
  \includegraphics[width=0.8\textwidth]{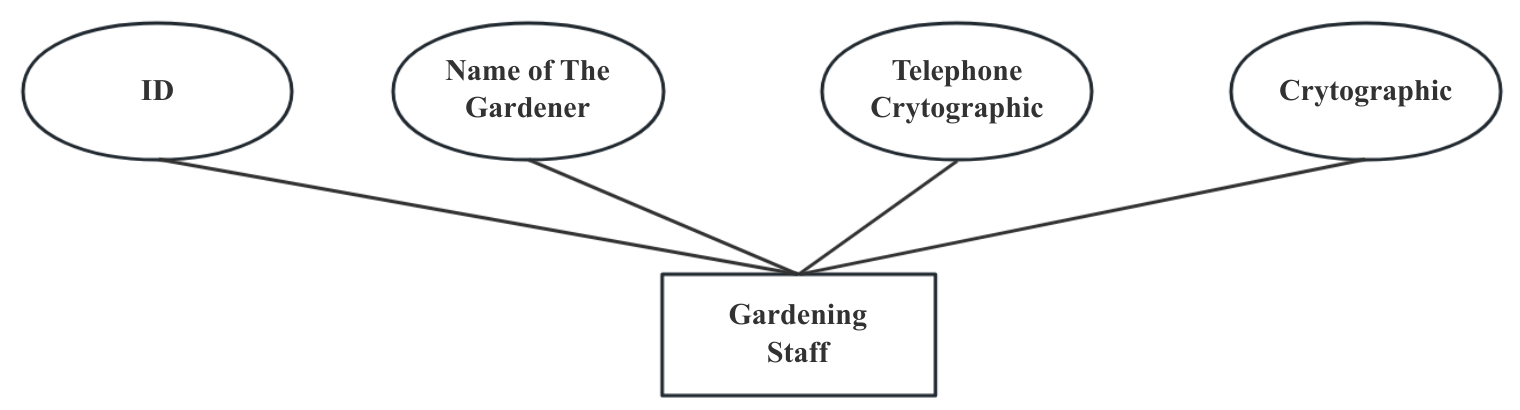}%width可以修改宽度
  \caption{Tree Management Entity Attribute Map}
   \label{fig:8}
\end{figure}

(4) \textbf{Gardener Personnel Entities}: The Gardener Personnel entity usually covers the Gardener Personnel Name, Password, Contact Number, and ID, as shown in the Gardener Personnel Entity Attribute Diagram(See Figure\ref{fig:8}). 

\begin{figure}[H] % [H] 表示强制当前位置
  \centering
  \includegraphics[width=0.8\textwidth]{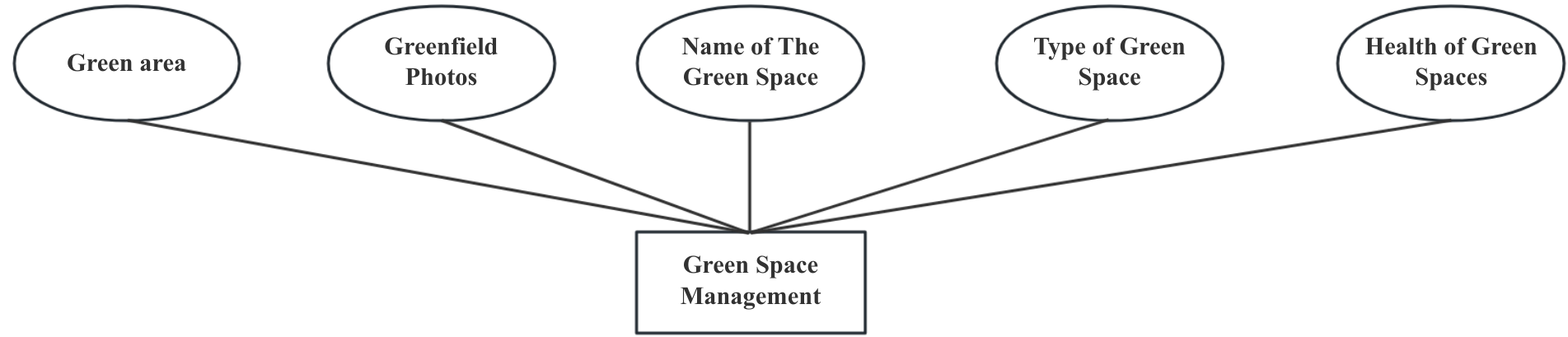}%width可以修改宽度
  \caption{Gardener Personnel Entity Attributes}
   \label{fig:9}
\end{figure}

(5) \textbf{Greenfield Entities}: The green space management entity usually stores attributes such as green space size, type, name of the green space, photo of the green space, health status of the green space, area of the green space, location of the green space, and so on(See Figure\ref{fig:9}).

\begin{figure}[H] % [H] 表示强制当前位置
  \centering
  \includegraphics[width=0.8\textwidth]{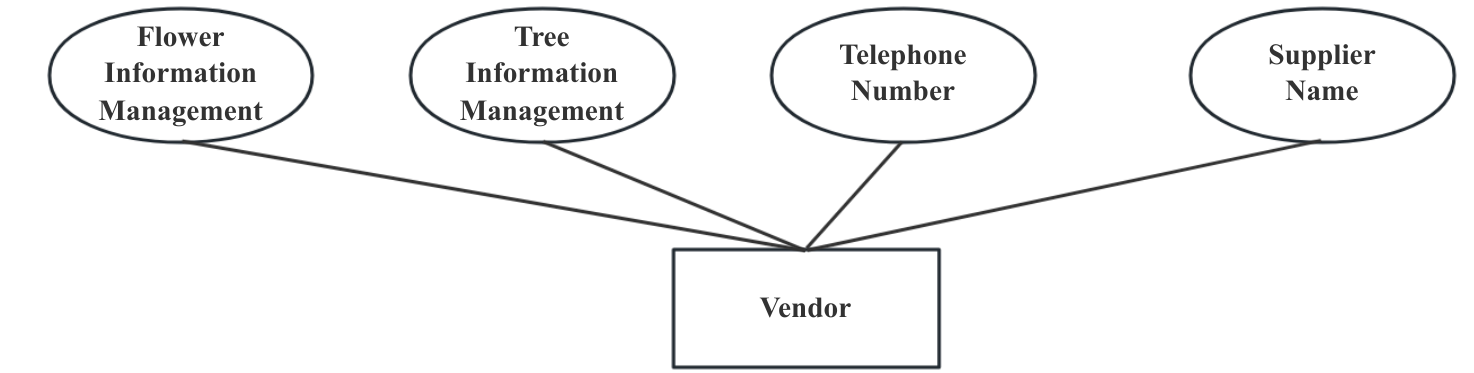}%width可以修改宽度
  \caption{Supplier Entity Attribute Diagram}
   \label{fig:10}
\end{figure}

(6) \textbf{Supplier Entities}: The supplier entity usually covers attributes such as flower type information, tree type information, contact phone number, supplier name, etc., and generates connection relationships with other entities(See Figure\ref{fig:10}). 

\begin{figure}[H] % [H] 表示强制当前位置
  \centering
  \includegraphics[width=0.8\textwidth]{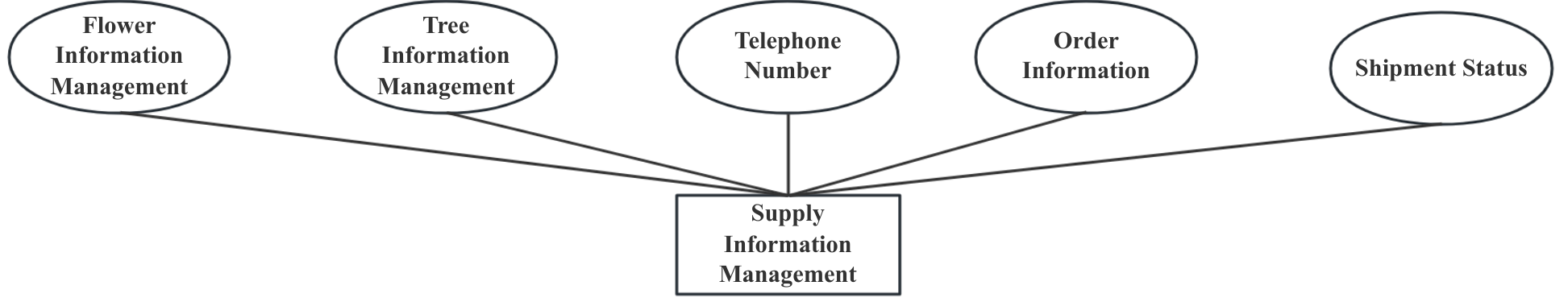}%width可以修改宽度
  \caption{Supplier Information Management Entity Attribute Diagram} \label{fig:11}
\end{figure}

(7) \textbf{Supply Information Management Entity}: Supply information management usually covers order information, tree information management, flower information management, contact phone number, shipping status, and other attributes. For the supply information management entity attributes of the contact chart (See Figure\ref{fig:11}).

\begin{figure}[H] % [H] 表示强制当前位置
  \centering
  \includegraphics[width=0.8\textwidth]{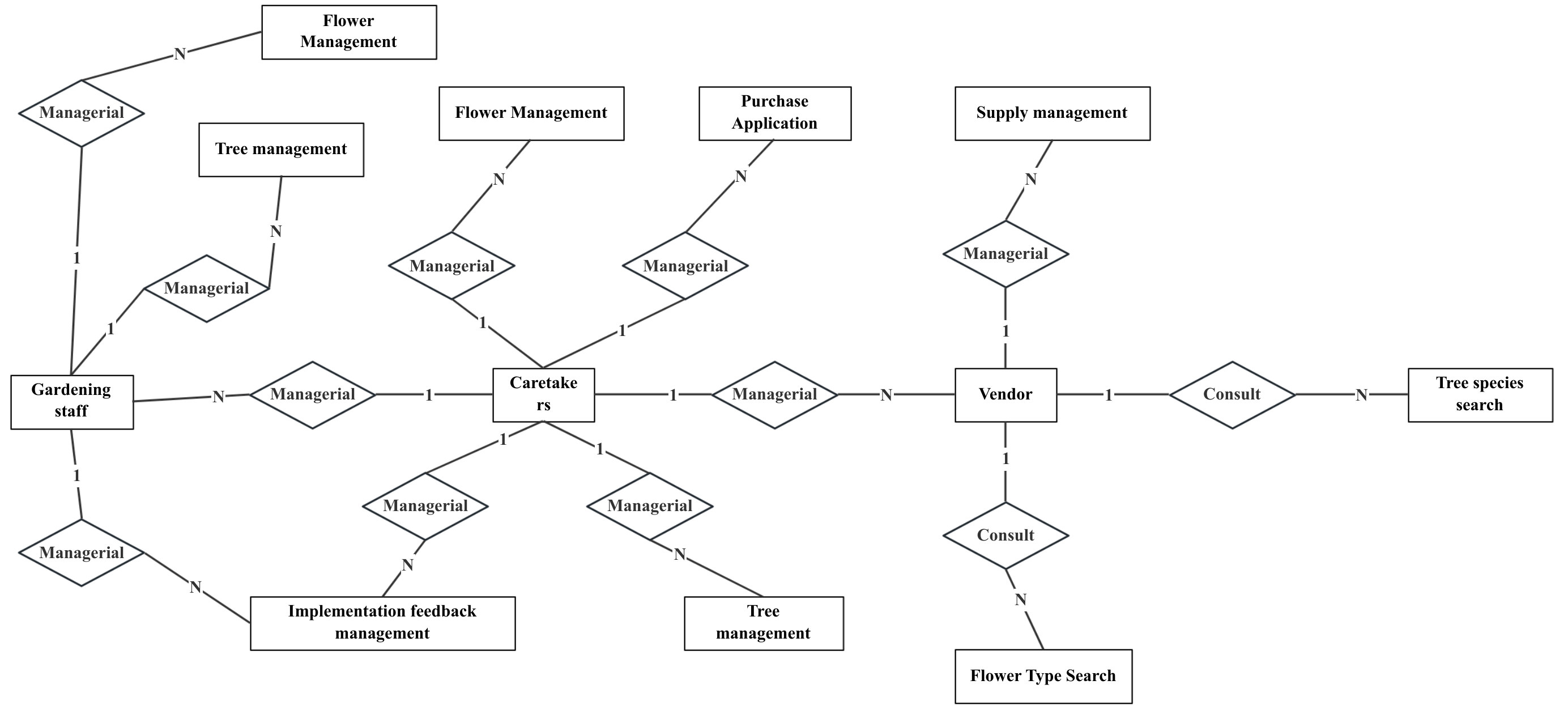}%width可以修改宽度
  \caption{System E-R Diagram}
   \label{fig:12}
\end{figure}

(8) The System E-R diagram mainly describes the relationship between various entities (See Figure\ref{fig:12}).

\subsubsection{Logical Design}
The administrator table stores information about the system administrator, including the administrator’s basic information, application records, and approval status. The administrator is responsible for the overall privilege management and assignment of the system. The design of the administrator table is shown in Table \ref{tab:1} : 
\begin{table}[H]
  \centering  % 表格居中
  \caption{Administrator Table}  % 表下正中间
  \label{tab:1}  % 引用标签
\begin{tabularx}{\textwidth}{l l X l l l }%%管理员表
 \toprule
  \textbf{Listings} & \textbf{Data Type} & \textbf{Whether or Not It Is Empty} & \textbf{Primary Key} & \textbf{Additive} & \textbf{Note}\\
  \midrule
  LDY\_Id  & Bigint(20) & Clogged & Be & Be & primary key \\
  LDY\_Username & Varchar(50) & Clogged & Clogged & Clogged & Administrator User Name \\
  LDY\_Password & Varchar(128) & Clogged & Clogged & Clogged & Administrator Password \\
  LDY\_Email & Varchar(100) & Be & Clogged & Clogged & Administrator Email \\
  \bottomrule
\end{tabularx}
\end{table}
The gardener personnel Table stores basic information about the gardener, including their name, contact information, and the area they are responsible for. The gardener is the core staff member accountable for the daily maintenance work in the Happy City Greening Management System. The design of the gardener personnel table is shown in Table \ref{tab:2} :
\begin{table}[H]
  \centering  % 表格居中
  \caption{Gardener Personnel Table}  % 表下正中间
  \label{tab:2}  % 引用标签
\begin{tabularx}{\textwidth}{l l X l l l}%园丁人员表
  \hline
  \textbf{Listings} & \textbf{Data Type} & \textbf{Whether or Not It Is Empty} & \textbf{Primary Key} & \textbf{Additive} & \textbf{Note}\\
  \hline
  LDY\_Id  & Bigint(20) & Clogged & Be & Be & Primary Key \\
  LDY\_Name & Varchar(50) & Clogged & Clogged & Clogged & Gardener's Name \\
  LDY\_Phone & Varchar(20) & Clogged & Clogged & Clogged & Gardener Contact
  Number \\
  LDY\_Email & Varchar(100) & Be & Clogged & Clogged & Gardeners Email \\
  LDY\_Hire\_Date & Varchar(24) & Clogged & Clogged & Clogged & Entry Date \\
  LDY\_Responsible\_Area & Varchar(128) & Be & Clogged & Clogged & Area of Responsibility \\
  \hline
\end{tabularx}
\end{table}
A flower table stores information about various types of flowers, including their names, types, planting quantities, planting locations, and maintenance requirements. The flower table is an integral part of the Happy Urban Greening Management System, which records detailed information about flowers. The design of the flower table is shown in Table \ref{tab:3} :
\begin{table}[H]
  \centering  % 表格居中
  \caption{Flower Table}  % 表下正中间
  \label{tab:3}  % 引用标签
\begin{tabularx}{\textwidth}{l l X l l l}%花卉表
  \hline
  \textbf{Listings} & \textbf{Data Type} & \textbf{Whether or Not It Is Empty} & \textbf{Primary Key} & \textbf{Additive} & \textbf{Note}\\
  \hline
  LDY\_Id  & Bigint(20) & Clogged & Be & Be & Primary Key \\
  LDY\_Name & Varchar(50) & Clogged & Clogged & Clogged & Gardener's Name \\
  LDY\_Category & Varchar(20) & Clogged & Clogged & Clogged & Types of Flowers\\
  LDY\_Quantity & Int(11) & Be & Clogged & Clogged & Number of Olants \\
  LDY\_Planting\_Area & Varchar(128) & Clogged & Clogged & Clogged & Planting Location \\
  LDY\_Maintenance\_Note & Varchar(256) & Be & Clogged &Clogged & Requirement\\
  \hline
\end{tabularx}
\end{table}
The tree table stores information about the various types of trees in the table, including the name, type, number of trees planted, planting location, and maintenance requirements. The tree table is an essential part of the Happy City Greening Management System, which records detailed information about trees. The design of the tree table is shown in Table \ref{tab:4} :
\begin{table}[H]
  \centering  % 表格居中
  \caption{Tree Table}  % 表下正中间
  \label{tab:4}  % 引用标签
\begin{tabularx}{\textwidth}{l l X l l l}%树木表
  \hline
  \textbf{Listings} & \textbf{Data Type} & \textbf{Whether or Not It Is Empty} & \textbf{Primary Key} & \textbf{Additive} & \textbf{Note}\\
  \hline
  LDY\_Id  & Bigint(20) & Clogged & Be & Be & Primary Key \\
  LDY\_Name & Varchar(50) & Clogged & Clogged & Clogged & Gardener's Name \\
  LDY\_Category & Varchar(20) & Clogged & Clogged & Clogged & Types of Flowers\\
  LDY\_Quantity & Int(11) & Be & Clogged & Clogged & NumBer of Plants \\
  LDY\_Planting\_Area & Varchar(128) & Clogged & Clogged & Clogged & Planting Location \\
  LDY\_Maintenance\_Note & Varchar(256) & Be & Clogged &Clogged & Requirement\\
  \hline
\end{tabularx}
\end{table}
The green space table records information about the distribution, area, person in charge, and maintenance schedule of green spaces in urban greening. The green space table is an integral part of the Happy Urban Greening Management System, which manages and maintains the green space resources in urban greening. The design of the green space table is shown in Table \ref{tab:5} :
\begin{table}[H]
  \centering  % 表格居中
  \caption{Green Space Table}  % 表下正中间
  \label{tab:5}  % 引用标签
\begin{tabularx}{\textwidth}{l X X X l X}%绿地表
  \hline
  \textbf{Listings} & \textbf{Data Type} & \textbf{Whether or Not It Is Empty} & \textbf{Primary Key} & \textbf{Additive} & \textbf{Note}\\
  \hline
  LDY\_Id  & Bigint(20) & Clogged & Be & Be & Primary Key \\
  LDY\_Name & Varchar(50) & Clogged & Clogged & Clogged & NumBer of The Green Space\\
  LDY\_Area & Decimal(10,2) & Clogged & Clogged & Clogged & Area of Green Space \\
 LDY\_Responsibility\_Gardener\_Id & Bigint(20) & Be & Clogged & Clogged & Responsible Gardener ID \\
  LDY\_Planting\_Area & Varchar(128) & Clogged & Clogged & Clogged & Planting Location \\
  LDY\_Maintenance\_Note & Varchar(256) & Be & Clogged & Clogged & Requirement\\
  \hline
\end{tabularx}
\end{table}
The purchase application form includes information such as the name of the purchased resource, the purchased quantity, the purchased price, and the delivery status. The purchase application form is an important part of the Happy City Green Management System, which records information on the purchased resources to facilitate unified management and improve supply efficiency. The design of the purchase application form is shown in Table \ref{tab:6} :
\begin{table}[htpb]
  \centering  % 表格居中
  \caption{Purchase Application}  % 表下正中间
  \label{tab:6}  % 引用标签
\begin{tabularx}{\textwidth}{l X X X l X}%购买表
  \hline
  \textbf{Listings} & \textbf{Data Type} & \textbf{Whether or Not It Is Empty} & \textbf{Primary Key} & \textbf{Additive} & \textbf{Note}\\
  \hline
  LDY\_Id  & Bigint(20) & Clogged & Be & Be & Primary Key \\
  LDY\_Name & Varchar(50) & Clogged & Clogged & Clogged & Purchase Resource Name\\
 LDY\_NumBer & Varchar(128) & Clogged & Clogged & Clogged & Quantity Purchased \\
  LDY\_Price & Bigint(20) & Clogged & Clogged & Clogged & Phrchase Price of Resources \\
  LDY\_Status & Varchar(24) & Be & Clogged & Clogged & Shipment Status\\
  \hline
\end{tabularx}
\end{table}

The supplier's basic information table is used to store the supplier's basic information, which includes the supplier's phone number, supplier name, contact number, and supply information management. As shown in Table \ref{tab:7} :
\begin{table}[H]
  \centering  % 表格居中
  \caption{Supplier Basic Information}  % 表下正中间
  \label{tab:7}  % 引用标签
\begin{tabularx}{\textwidth}{l X X X l X}%供货信息表
  \hline
  \textbf{Listings} & \textbf{Data Type} & \textbf{Whether or Not It Is Empty} & \textbf{Primary Key} & \textbf{Additive} & \textbf{Note}\\
  \hline
  LDY\_Id  & Bigint(20) & Clogged & Be & Be & Primary Key \\
  LDY\_Name & Varchar(50) & Clogged & Clogged & Clogged & Supplier Name\\
 LDY\_Phone & Varchar(20) & Clogged & Clogged & Clogged & Supplier Contact Number\\
  LDY\_Information & Varchar(100) & Be & Clogged & Clogged & Supply
  Information Management \\
  \hline
\end{tabularx}
\end{table}

\section{Implementation}
\subsection{Administrator Module}
\subsubsection{Tree Management}
After taking a photo of a tree and filling in a description of the problem(See the Listing\ref{code1}), the caretaker calls the back-end interface (e.g., POST /api/trees/{id}/report) and executes HealthCheckScheduler.dailyCheck() daily, which scans all the tree data to detect the health status of the tree as well as to update the tree species.

\begin{lstlisting}[language=Java,caption={Basic Functions in Tree Management},label={code1}]
// TreeController.java
@RestController
@RequestMapping("/api/trees")
@Api(tags = "Tree Management API")
public class TreeController {
    
    @Autowired
    private TreeService treeService;
        
    @PostMapping
    @PreAuthorize("hasRole('ADMIN')")
    public ApiResult addTree(@Valid @RequestBody TreeAddDTO dto) {
        treeService.addTree(dto);
        return ApiResult.success();
    }    
   
    @GetMapping("/{id}")
    public ApiResult<TreeDetailVO> getTreeDetail(@PathVariable Long id) {
        return ApiResult.success(treeService.getDetail(id));
    }   
    
    @PostMapping("/{id}/image")
    public ApiResult uploadImage(
        @PathVariable Long id, 
        @RequestParam("file") MultipartFile file
    ) {
        String imageUrl = fileService.uploadTreeImage(file);
        treeService.updateTreeImage(id, imageUrl);
        return ApiResult.success(imageUrl);
    }
}
\end{lstlisting}

\subsubsection{Tree Species Management}
First, front-end request(See the Listing\ref{code2}): Users enter keywords (e.g., "pine") or select family filter conditions (e.g., "pine") in the front-end interface, and click the query button. Secondly, HTTP request: The front-end sends a GET request to the back-end interface /api/species, carrying the parameters keyword=Pine\&family=Pine\&page=1\&pageSize=10. Parameter validation: The back-end receives the parameters through TreeSpeciesQueryDTO, and finally, the parameters are automatically validated for legality. Parameter validation: The backend receives the parameters through TreeSpeciesQueryDTO, and at the end, the parameters are automatically validated.

\begin{lstlisting}[language=Java,caption={Basic Management Functions for Tree Species},label={code2}]
@RestController
@RequestMapping("/api/species")
@Validated
public class TreeSpeciesController {
   
    @Autowired
    private TreeSpeciesService treeSpeciesService;  
    @GetMapping
    public ResponseEntity<PageResponse<TreeSpecies>> searchSpecies(
            @Valid TreeSpeciesQueryDTO queryDTO, 
            BindingResult bindingResult) {
        
        if (bindingResult.hasErrors()) {
            throw new InvalidParameterException(bindingResult.getFieldError().getDefaultMessage());
        }
        return ResponseEntity.ok(treeSpeciesService.search(queryDTO));
    }
}
}
\end{lstlisting}
\subsubsection{Flower Management}
Firstly, front-end request(See the Listing\ref{code-1}): The administrator fills in flower information (name, family, flowering period, description, etc.) in the front-end and uploads flower images, clicks the submit button. Secondly, HTTP request: The front-end sends a POST request to /api/flowers, the request body is in multipart/form-data format, containing form data and image files. Finally, permission verification: Spring Security verifies the user's permissions via @PreAuthorize("hasRole('ADMIN')").

\begin{lstlisting}[language=Java,caption={Basic Functions in Flower Management Functions},label={code-1}]
@Service
public class FlowerService {

    @Value("${upload.directory}")
    private String uploadDirectory; 
    public String saveImage(MultipartFile image) throws IOException {
        
        if (image.isEmpty()) throw new IllegalArgumentException("Not Empty");
        String fileName = UUID.randomUUID() + "_" + image.getOriginalFilename();
        
        Path filePath = Paths.get(uploadDirectory, fileName);
        Files.createDirectories(filePath.getParent());
        Files.write(filePath, image.getBytes());

        return "/uploads/" + fileName; 
    }
\end{lstlisting}

\subsubsection{Management of Floral Species}
First of all, front-end form submission: The administrator fills in flower information (name, type) in the front-end, selects the image file, and clicks the "Submit" button. HTTP request(See the Listing\ref{code3}): The front-end sends a POST request to /api/flower-species in the format of multipart/form-data, containing form fields and image files. Data, containing form fields and image files. Permission validation: Spring Security validates the user role via @PreAuthorize("hasRole('ADMIN')") and returns 403 Forbidden for non-administrators.

\begin{lstlisting}[language=Java,caption={Basic Flower Species Management Functions},label={code3}]
    @Service
@RequiredArgsConstructor
public class FlowerSpeciesService {
    private final FlowerSpeciesMapper speciesMapper.
    private final FileStorageService fileStorageService; 
    @Transactional
    public FlowerSpecies createSpecies(FlowerSpeciesCreateDTO dto, MultipartFile imageFile) {      
        if (speciesMapper.exists(Wrappers.<FlowerSpecies> lambdaQuery())
                .eq(FlowerSpecies::getCommonName, dto.getCommonName())) {
            throw new BusinessException(ErrorCode.SPECIES_NAME_EXISTS);
        }
        String imageUrl = null;     
        if (imageFile ! = null && !imageFile.isEmpty()) {
            imageUrl =  fileStorageService.storeFlowerImage(imageFile);
        }

\end{lstlisting}
\subsubsection{Green Space Management}
Firstly, front-end request: Administrator fills in green-space information (name, area, latitude, longitude, type, status) in the front-end, uploads multiple images, and clicks submit, and secondly, HTTP request(See the Lisitng\ref{code4}): Sends POST /api/green-spaces, The request is in the format of multipart/form-data, and contains form data and image files. Finally, permission checking: Validate the administrator permission by @PreAuthorize("hasRole('ADMIN')"). 

\begin{lstlisting}[language=Java,caption={Basic Management Functions for Green Spaces},label={code4}]

@PostMapping(consumes = MediaType.MULTIPART_FORM_DATA_VALUE)
public ResponseEntity< GreenSpace>  addGreenSpace(   
    @Validated @ModelAttribute GreenSpaceCreateDTO dto,
    @RequestPart List<MultipartFile> images
) {    
    if (greenSpaceMapper.existsByName(dto.getName())) {
        throw new BusinessException("The name of the green space already exists");;
    }
    List<String> imageUrls =  images.stream()
        .map(fileStorageService::storeGreenSpaceImage)
        .collect(Collectors.toList());
        
    GreenSpace space = GreenSpace.builder()
        .name(dto.getName())
        .area(dto.getArea())
        .location(new Point(dto.getLongitude(), dto.getLatitude()))
        .imageUrls(JSON.toJSONString(imageUrls))
    greenSpaceMapper.insert(space);   
    return ResponseEntity.status(HttpStatus.CREATED).body(space);
}

\end{lstlisting}
\subsubsection{Management of Conservation Plans}
First, update the task status to progress by calling /api/tasks/{taskId}/accept. Upload before and after photos to OSS by calling FileService.uploadImage(). Fill in the maintenance log (e.g., medication dosage, operation duration), ensure data consistency through @Transactional. Call /api/tasks/{taskId}/complete after administrator review, update task status to COMPLETED(See the Listing\ref{code5}). 

\begin{lstlisting}[language=Java,caption={Basic Management Functions For Conservation Programmes},label={code5}]
    @PostMapping("/tasks/{taskId}/complete")
@PreAuthorize("hasRole('WORKER')")
public ApiResult completeTask(@PathVariable Long taskId, @RequestBody TaskCompletionDTO dto) {
    maintenanceService.completeTask(taskId, dto.getPhotos(), dto.getNotes());
    treeService.reEvaluateHealth(taskId); 
    return ApiResult.success();
}
\end{lstlisting} 

\subsection{Gardener Personnel Module}
\subsubsection{Feedback Management}
After completing the maintenance tasks, the gardener personnel can submit executive feedback through the system, including completing the functions, problems encountered, and solutions. The system supports the input submission, viewing, and replying functions, which facilitate the communication between gardeners and administrators. The gardener personnel can also adjust the work method and process in time according to the feedback content to improve the conservation efficiency and quality(See the Listing\ref{code-2}).

\begin{lstlisting}[language=Java,caption={Basic Feedback Management Function},label={code-2}]
@Service
public class FeedbackService {
    @Autowired
    private FeedbackRepository feedbackRepository;

    @Transactional
    public Feedback createFeedback(FeedbackCreateRequest request, Long submitterId) {
        Feedback feedback = new Feedback();
        feedback.setTitle(request.getTitle());
        feedback.setContent(request.getContent());
        feedback.setSubmitterId(submitterId);
        return feedbackRepository.save(feedback);
    }
    public List<FeedbackResponse> getFeedbacks(UserPrincipal user) {
        List<Feedback> feedbacks;
        if (user.hasRole("ADMIN")) {
            feedbacks = feedbackRepository.findAll();
        } else {
            feedbacks = feedbackRepository.findBySubmitterId(user.getId());
        }
        return feedbacks.stream().map(this::convertToResponse).toList();
    }
}

\end{lstlisting}

\subsection{Supplier Module}
\subsubsection{Supply Management}
First of all, the front-end operation supplier fills out the form (user name, real name, cell phone number, product name, quantity) and clicks submit(See the Listing\ref{code6}). At the same time HTTP request is sent to send a POST /api/supplies, the request body for JSON format for the supply information. Next, repeat the submission check to check whether the same cell phone number is pending (pending applications. Check the initialization audit status as pending and set the creation time. Call supplyMapper.insert() to insert into the database table supply\_manage. Finally, the response returns 201 Created and the newly created supply order information. 

\begin{lstlisting}[language=Java,caption={Basic Management Functions for Suppliers},label={code6}]
@PostMapping
public ResponseEntity<Supply> submitSupply(@Validated 
@RequestBody SupplySubmitDTO dto) {
    if (supplyMapper.existsByPhoneAndPending(dto.getPhone())) {
        throw new BusinessException("There is already a pending application for this cell phone number.");        
    }
    supplyMapper.insert(supply);
    return ResponseEntity.status(HttpStatus.CREATED).body(supply);   
}
\end{lstlisting}

\subsubsection{Flower Species Search}
First, front-end input conditions(See the Listing\ref{code7}): The user enters the search keywords (e.g., name "Rose"), selects the family (e.g., "Rosaceae"), and clicks the "Search" button. Next, HTTP request: Send a GET request to /api/flower-species? name=rose\&family=Rose family. Finally, parameter parsing: The backend receives the query parameters via FlowerSpeciesQueryDTO and automatically checks the parameter format (e.g., field lengths). 

\begin{lstlisting}[language=Java,caption={Basic Functions of Flower Type Search},label={code7}]
    @PostMapping("/search")
@ApiOperation("Advanced Feature Retrieval")
public Result<Page<FlowerSpecies>>  advancedSearch(
@RequestBody SpeciesSearchDTO dto,
@PageableDefault(size = 10) Pageable pageable) {
        return Result.success(speciesService.advancedSearch(dto, pageable));
    }
@GetMapping("/family/{family}")
@ApiOperation("Search by Family")
    public Result<List<FlowerSpecies>>  getByFamily(
@PathVariable String family) {
        return Result.success(speciesMapper.selectList(
            Wrappers.<FlowerSpecies> query().eq("family", family)
        ));
    }
\end{lstlisting}
\subsubsection{Tree Species Search}
The first front-end operation is performed(See the Listing\ref{code8}): The user clicks on a tree species entry. Send a GET request via HTTP to /api/tree-species/1. Secondly, data query: Call treeSpeciesMapper.selectById(1) to query the database. If it doesn't exist, throw BusinessException("Specified tree species not found"). Final data conversion: Convert the TreeSpecies entity to TreeSpeciesDetailVO and parse the distribution field. 
\begin{lstlisting}[language=Java,caption={Basic Functions of Tree Species Search},label={code8}]
@Service
@RequiredArgsConstructor
public class TreeSpeciesService {
    private final TreeSpeciesMapper treeSpeciesMapper;
    public PageResult< TreeSpeciesDetailVO>  searchSpecies(TreeSpeciesQueryDTO query) 
    public TreeSpeciesDetailVO getSpeciesDetail(Integer id) {
        TreeSpecies species = treeSpeciesMapper.selectById(id);
        if (species == null) {
            throw new BusinessException("The specified tree species was not found");.
        }
        return convertToDetailVO(species).
    }
    }
}
\end{lstlisting}
\section{Evaluation}
    
Testing can find the problem areas of the system. In the process of system development, to ensure the quality of the code, need to carry out various forms of testing of the system, which can help to find potential problems in the system, such as anomalies, loopholes, and so on \cite {v, grossmanPracticalVerificationSmart2024}. Testing can guarantee delivery quality; testing can effectively improve the stability and reliability of the system to ensure that the system can be successfully delivered online and ensure good operation. Testing can improve user satisfaction. Testing can check the user's perspective and further optimize their experience to enhance user satisfaction\cite{w, liDemoEnhancingSmart2024}.

\subsection{Use Cases}
The system's testing was focused on the following modules: Tree Management Module, Flower Management Module, Green Space Management Module, and Order Request Module. The testing covered functionality testing, which included adding, inserting, deleting, querying, and applying to ensure that each module worked properly under different functions\cite{priftiSmartContractVulnerability2024}..

\begin{itemize}

\item[(1).] Check if the module interface is loaded correctly.
\item[(2).] Test that the add, query, modify, and delete functions work as expected.
\item[(3).] Input error data to verify that the system can handle the exception correctly.
\item[(4).]Check data persistence to ensure consistent front-end and back-end interactions and database operations.

\end{itemize}

Test case numbers are named after module functions and operation types, which clarify the test objectives. The accuracy and stability of the module functions are verified by testing different conditions and input data\cite{x, weiSurveyQualityAssurance2024}. This test focuses on adding, inserting, deleting, and querying tree information in Tree Management, with test cases to show that the module functions as expected\cite{zhangInferringLikelyCountingrelated2025}(See the Table\ref{TC:1}):

\begin{longtable}{ l p{3cm} p{3cm} p{3cm} p{2.5cm} } % 手动设定列宽
  \caption{Tree Management Test Case Table} \label{TC:1} \\
  \hline
\textbf{Serial Number} & \textbf{Test Case Description} & \textbf{Input Data} & \textbf{Projected Output} & \textbf{Actual Output}\\
  \hline
  \endfirsthead % 分页表头
    \multicolumn{5}{c}{\textmd{Table \thetable~ (Continued): Tree Management Test Case Table}} \\ % 续表的标题行
  \hline
\textbf{Serial Number} & \textbf{Test Case Description} & \textbf{Input Data} & \textbf{Projected Output} & \textbf{Actual Output}\\
  \hline
  \endhead % 后续页重复表头
  
  \hline
  \endfoot % 表格底部
  TC\_Add True\_01 & 1. Add information about poplar trees
  
  2. Query the existence of poplar tree information & Various Trees of The Genus Populus & Tree information has been success fully added & Pass a Test\\
  TC\_SeleteTrue\_02& 1. Enter acacia
2. Query the existence of acacia information
 & Locust Tree (Sophora Japonica) & Relevant tree information found & Pass a Test\\
 TC\_Alter True\_03  & 1. Enter information about the willow tree

2. Modify information on willow trees to acacia trees
 & Willow Tree & The tree information has been successfully updated & Pass a Test\\
  TC\_Delete True\_04 & 1. Enter information about the acacia tree
  
2. Delete the information on acacia trees 
& Locust Tree(Sophora Japonica) & The tree information has been successfully removed & Pass a Test\\
  TC\_Add Plan\_05 &1.Add information about the popular program
  
2.Query the existence of popular program information& The Poplar Project & Add successfully & Pass a Test\\
  \hline
\end{longtable}

This test focuses on adding, inserting, deleting, and querying the flower information of the Flower Manager, with test cases to show that the module functions as expected(See the Table\ref{TC:2}):
\begin{longtable}{ l p{3cm} p{3cm} p{3cm} p{2.5cm} } % 手动设定列宽
  \caption{Flower Management Test Case Table} \label{TC:2} \\
  \hline
\textbf{Serial Number} & \textbf{Test Case Description} & \textbf{Input Data} & \textbf{Projected Output} & \textbf{Actual Output}\\
  \hline
  \endfirsthead % 分页表头
    \multicolumn{5}{c}{\textmd{Table \thetable~ (Continued): Flower Management Test Case Table}} \\ % 续表的标题行
  \hline
\textbf{Serial Number} & \textbf{Test Case Description} & \textbf{Input Data} & \textbf{Projected Output} & \textbf{Actual Output}\\
  \hline
  \endhead % 后续页重复表头
  
  \hline
  \endfoot % 表格底部
  
  % ---------- 表格内容 ----------
  TC\_Add Flower\_01  & 
  1. Add information aboutroses \ newline
  2. Query the existence of rose information & 
  Rose & 
  Flower information added successfully & 
  Pass a Test \\
  
  TC\_Selete Flower\_02 & 
  1. Inquire about moonflowers & 
  Chinese Rose (Rosa Sinensis) & 
  Successful query & 
  Pass a Test \\
  
  TC\_Alter Flower\_03  & 
  1. Add information about the moonflower\newline
  2. Modify the information to rose flower & 
  Chinese Rose (Rosa Sinensis) & 
  Moonflower information updated successfully & 
  Pass a Test \\
  
  TC\_Delete Flower\_04 & 
  1. Search for chrysanthemums information\newline
  2. Delete chrysanthemums information & 
  Chrysanthemum Deleted Successfully & 
  Chrysanthemum has been successfully deleted & 
  Pass a Test \\
  
  TC\_Add Flower Plan\_05 & 
  1. Add a rose care program\newline
  2. Check if program was added successfully & 
  Rose Care Program & 
  Rose care program on the calendar & 
  Pass a Test \\
  \hline
\end{longtable}

This test focuses on adding, inserting, deleting, and querying the green space information of the Green Space Management to show through test cases that the module functions as expected(See the Table\ref{TC:3}):

\begin{longtable}{ l p{3cm} p{3cm} p{3cm} p{2.5cm} } % 手动设定列宽  \centering  % 表格居中
  \caption{Green Space Management Test Case Table}  % 表下正中间
    \label{TC:3} \\
  \hline
\textbf{Serial Number} & \textbf{Test Case Description} & \textbf{Input Data} & \textbf{Projected Output} & \textbf{Actual Output}\\
  \hline
  \endfirsthead % 分页表头
  \multicolumn{5}{c}{\textmd{Table \thetable~ (Continued): Green Space Management Test Case Table}} \\ % 续表的标题行
  \toprule
\textbf{Serial Number} & \textbf{Test Case Description} & \textbf{Input Data} & \textbf{Projected Output} & \textbf{Actual Output}\\
  \hline
  \endhead % 后续页重复表头
  \hline
  \endfoot % 表格底部
  TC\_Add Green\_01  &1. Add information about large lawns& Large Lawn& Large lawn information added & Pass a Test\\
  TC\_Selete Green\_02&1. Search for information on large lawns
 & Large Lawn &Successful search for large lawn information& Pass a Test\\
 TC\_Alter Green\_03  & 1. Add information on wasteland
 
 2. Revision of wasteland information to green belt
 & Uncultivated Land &Wasteland information modified successfully& Pass a Test\\
TC\_Delete Green\_04  & 1. Enquire about the Green belt
  2. Deletion of greenbelt information & Green Belt& CGreen belt Deletion Successful& Pass a Test\\

TC\_Add Green Plan\_05 &1. Add a lawn care program

2. Check if the lawn care program was added successfully & Maintenance
program developed and associated with lawn success & Maintenance program developed and associated with lawn success & Pass a Test\\
  \hline
\end{longtable}

This test focuses on adding, inserting, deleting, and querying the order information of the order request management and uses test cases to show that the module's functionality works as expected(See the Table\ref{TC:4}):

\begin{longtable}{ l p{3cm} p{3cm} p{3cm} p{2.5cm} } % 手动设定列宽
  \caption{Order Request Test Form} 
  \label{TC:4} \\
  \hline
\textbf{Serial Number} & \textbf{Test Case Description} & \textbf{Input Data} & \textbf{Projected Output} & \textbf{Actual Output}\\
  \hline
  \endfirsthead % 分页表头
  \multicolumn{5}{c}{\textmd{Table \thetable~ (Continued): Order Request Test Form}} \\ % 续表的标题行
  \toprule
\textbf{Serial Number} & \textbf{Test Case Description} & \textbf{Input Data} & \textbf{Projected Output} & \textbf{Actual Output}\\
  \hline
  \endhead % 后续页重复表头
  \hline
  \endfoot % 表格底部
  
  % ---------- 表格内容 ----------
  TC\_Add order\_01   &1. Add information about the order

2. Query whether the order information exists& Add Order& Purchase order
added successfully& Pass a Test\\
 TC\_Selete order\_02&1. Query order  &Check Order Details &Return to
Purchase Order Details& Pass a Test\\
 TC\_Alter order\_03& 1. Check the information on the order
 
2. Modify order information
 & Modify Order Information&Purchase Order Information Modified Successfully& Pass a Test\\
  TC\_Delete order\_04   & 1. Check the information on the order
  
2. Delete order information& Delete Purchase Order& Purchase order
deleted successfully& Pass a Test\\
  TC\_Realize order\_05&1. Check the information on the order feedback
  
2. Fulfillment of Order Feedback & Order Fulfillment and Feedback& Feedback
Recorded Successfully Purchase Order Updated& Pass a Test\\
  \hline
\end{longtable}

\subsection{Discussion}
The gardener personnel management test case demonstrates that the system accurately performs operations to add, modify, and delete gardener information while providing clear notifications for illegal or unauthorized actions, thereby confirming its reliability. In the tree management test case, the system efficiently manages the addition, deletion, modification, and verification of tree information in conjunction with the execution of maintenance plans, thus emphasizing its functionality. The flower management test case showcases adequate supervision of flower information and provides a detailed list of flower species, facilitating the development of smooth maintenance plans. In the green space management test case, the system comprehensively oversees green space information and successfully associates maintenance plans with relevant areas. Finally, the conservation plan management test case verifies that the system can flexibly create, modify, and delete conservation plans while precisely reflecting the outcomes after their execution. Overall, the system exhibits excellence across all functions, showcasing robust capabilities and effectively supporting urban greening management.  

\section{Conclusion}
The Spring Boot-architected Happy City Phytogeographical Governance Platform has been validated through comprehensive stress testing, demonstrating robust capabilities in coordinating multi-stakeholder operational elements, including horticultural workforce management, supplier ecosystems, procurement workflows, dendrological inventories, floricultural databases, greenspace cadastres, and maintenance scheduling optimization. This systemic integration substantially enhances the efficacy of municipal phytocentric governance by automating process orchestration. Leveraging Spring Boot’s convention-over-configuration paradigm, the solution achieves high-performance execution capabilities while maintaining architectural elasticity. The implementation leverages the extensive technological ecosystem offered by Spring, particularly its mechanisms for dependency injection and modular design paradigms. These features significantly contribute to the sustainability of the codebase and enhance the efficiency of iterative development processes. The platform incorporates an ergonomic human-machine interface engineered with cognitive load reduction principles, enabling municipal operators to execute complex photo management tasks through streamlined operational workflows. Our continuous improvement protocol employs agile DevOps practices, integrating IoT-enabled environmental sensing networks for real-time photometric monitoring and predictive maintenance analytics. Ongoing system optimization prioritizes fault-tolerant architecture enhancements through circuit breaker patterns and zero-downtime deployment strategies. Security hardening initiatives implement OAuth 2.0 authorization protocols and role-based access control matrices. Architectural innovations under exploration include cloud-native serverless architectures and containerized microservices, which enhance horizontal scalability and flexibility.

\section {Acknowledgments}
The author would like to express sincere gratitude to Prof. Fu Lifang for her
invaluable guidance and support during the foundational stages of this research.

\bibliographystyle{unsrt}  
\bibliography{references}  

@article{a,
  author   = {Wu, Xiaoyun and Yuan, Haodong},
  title    = {Online Examination Management System Based on SpringBoot},
  journal  = {Microcomputer Applications},
  year     = {2024},
  volume   = {40},
  number   = {11},
  pages    = {199--204},
  language = {Chinese}
}

@article{b,
  author   = {Jia, Wenqiang and Liu, Xin and Fu, Peng},
  title    = {Design and Implementation of Enterprise Record Management System Based on SpringBoot and Vue Framework},
  journal  = {Industrial Control Computer},
  year     = {2024},
  volume   = {37},
  number   = {10},
  pages    = {151--152},
  language = {Chinese}
}

@article{c,
  author   = {Xie, Zhenhua},
  title    = {Design of Educational Administration System Based on Vue.js and SpringBoot},
  journal  = {Computer and Information Technology},
  year     = {2024},
  volume   = {32},
  number   = {4},
  pages    = {95--97,101},
  language = {Chinese}
}

@book{d,
  author    = {Palacios, Miguel F.},
  title     = {SpringBoot 3.0 Cookbook: Proven Recipes for Building Modern and Robust Java Web Applications with SpringBoot},
  publisher = {Packt Publishing Limited},
  year      = {2024},
  month     = {Jul},
  day       = {12}
}

@book{e,
  author    = {Memiş, Ahmet},
  title     = {Mastering SpringBoot 3.0: A Comprehensive Guide to Building Scalable and Efficient Backend Systems with Java and Spring},
  publisher = {Packt Publishing Limited},
  year      = {2024},
  month     = {Jun},
  day       = {28}
}

@article{f,
  author   = {Yang, Deshun and Yang, Shuzhen},
  title    = {Textbook Management System Based on SpringBoot},
  journal  = {Computer Programming Skills \& Maintenance},
  year     = {2024},
  number   = {5},
  pages    = {100--103},
  language = {Chinese}
}

@article{g,
  author   = {Jiang, Shaohua and Chang, Xinghai and Gao, Yunfan and et al.},
  title    = {Design and Application of Integrated Sheep Farm Management System Software Based on SpringBoot and Vue Framework},
  journal  = {Journal of Domestic Animal Ecology},
  year     = {2024},
  volume   = {45},
  number   = {3},
  pages    = {55--62},
  language = {Chinese}
}

@article{h,
  author   = {Ding, Fujiang},
  title    = {Design and Implementation of Visitor Management System Based on SpringBoot and Vue Technology},
  journal  = {Computer Programming Skills \& Maintenance},
  year     = {2023},
  number   = {12},
  pages    = {82--87},
  language = {Chinese}
}

@article{i,
  author   = {Zhao, Jing and Wang, Yongle},
  title    = {Research on Intelligent Ecological Management System for Gardens Based on Wireless Networking Technology},
  journal  = {Modern Electronics Technique},
  year     = {2024},
  volume   = {47},
  number   = {20},
  pages    = {160--164},
  language = {Chinese}
}

@article{j,
  author   = {Pan, Xiaochen and Wang, Cheng and Wang, Jie},
  title    = {Research on Construction and Management Practices of Pollinator Gardens in the United States and Its Enlightenment to China},
  journal  = {Landscape Architecture},
  year     = {2023},
  volume   = {40},
  number   = {9},
  pages    = {116--122},
  language = {Chinese}
}

@book{l,
  author    = {Memiş, Ahmet},
  title     = {Mastering SpringBoot 3.0: A Comprehensive Guide to Building Scalable and Efficient Backend Systems with Java and Spring},
  year      = {2024},
  publisher = {Packt Publishing Limited},
  month     = {6}
}

@article{m,
  author   = {Wu, Xiaoyun and Yuan, Haodong},
  title    = {Online Examination Management System Based on SpringBoot},
  journal  = {Microcomputer Applications},
  year     = {2024},
  volume   = {40},
  number   = {11},
  pages    = {199--204},
  language = {Chinese},
  note     = {SpringBoot Technical Architecture Reference}
}

@article{n,
  author     = {Jiang, Shaohua and Chang, Xinghai and Gao, Yunfan and et al.},
  title      = {Design and Application of Integrated Livestock Farm Management System Software Based on SpringBoot and Vue Framework},
  journal    = {Journal of Domestic Animal Ecology},
  year       = {2024},
  volume     = {45},
  number     = {3},
  pages      = {55--62},
  language   = {Chinese}
}

@article{o,
  author   = {Zhao, Jing and Wang, Yongle},
  title    = {Research on Intelligent Ecological Management System for Gardens Based on Wireless Networking Technology},
  journal  = {Modern Electronics Technique},
  year     = {2024},
  volume   = {47},
  number   = {20},
  pages    = {160--164},
  language = {Chinese}
}

@misc{p,
  author       = {{Ministry of Natural Resources}},
  title        = {National Territorial Spatial Planning Implementation Monitoring Network Construction Work Plan (2023-2027)},
  year         = {2023},
  howpublished = {Policy Document [Z]},
  note         = {Department of Natural Resources, Bureau of Territorial Spatial Planning},
  langid       = {Chinese}
}

@inproceedings{q,
  author    = {Zhou, Yuanyuan and Tang, Zili and Zhang, Bo and et al.},
  title     = {Design and Implementation of Image Sample Management Database},
  booktitle = {Seventh Symposium on Novel Photoelectronic Detection Technology and Applications},
  year      = {2021},
  volume    = {11763},
  series    = {SPIE Proceedings},
  pages     = {117631H},
  doi       = {10.1117/12.2593258}
}

@techreport{r,
  author      = {{Shanghai Municipal Planning and Natural Resources Bureau}},
  title       = {CSPON Construction: Empowering Megacity Governance Through Full-Element Living Circle Intelligent Assessment},
  year        = {2025},
  type        = {Technical Report},
  institution = {Natural Resources Ministry},
  langid      = {Chinese}
}

@article{u,
  author  = {{United Nations}},
  title   = {The New Urban Agenda},
  year    = {2016},
  journal = {Habitat III Conference Report},
  pages   = {1--175},
  url     = {https://habitat3.org/the-new-urban-agenda/}
}

@article{v,
  author   = {Liu, Xianglong and Zeng, Li},
  title    = {Database Design and Implementation of Cinema System},
  journal  = {Computer Knowledge and Technology},
  year     = {2022},
  volume   = {18},
  number   = {06},
  pages    = {16--18},
  langid   = {Chinese}
}

@article{w,
  author  = {Cheng, Yuan and Chen, Chunhua and Zhu, Jingxian and et al.},
  title   = {Nuclear Emergency Rescue Drill Database Design and Implementation},
  journal = {Annals of Nuclear Energy},
  year    = {2022},
  volume  = {166},
  pages   = {108957},
  doi     = {10.1016/j.anucene.2021.108957}
}

@article{x,
  author     = {Wang, Fang},
  title      = {Design and Implementation of Informatized Educational Administration System},
  journal    = {Information Recording Materials},
  year       = {2021},
  volume     = {22},
  number     = {02},
  pages      = {223--225},
  langid     = {Chinese}
}

@article{y,
  author   = {Li, Cheng and Hu, Wuyin},
  title    = {Application of Java Programming Language in Computer Software Development},
  journal  = {Electronic Technology},
  year     = {2024},
  volume   = {53},
  number   = {03},
  pages    = {66--67},
  langid   = {Chinese}
}

@inproceedings{xiao2025parallelizing,
  title     = {Parallelizing Universal Atomic Swaps for $\{$Multi-Chain$\}$ Cryptocurrency Exchanges},
  author    = {Xiao, Danlei and Zhang, Chuan and Deng, Haotian and Liang, Jinwen and Wang, Licheng and Zhu, Liehuang},
  booktitle = {Proceedings of the 34th USENIX Security Symposium (USENIX Security 25)},
  pages     = {4073--4092},
  year      = {2025}
}

@inproceedings{aguilarSmartContractFamilies2024,
  title     = {Smart {{Contract Families}} in {{Solidity}}},
  booktitle = {Proceedings of the 34th {{International Conference}} on {{Collaborative Advances}} in {{Software}} and {{COmputiNg}} (CASCON)},
  author    = {Aguilar, Julio and Bak, Kacper and Boyle, Michael and Callens, Valerian and Gorzny, Jan},
  year      = 2024,
  pages     = {1--5}
}

@inproceedings{arceriSoundConstructionEVM2024,
  title     = {Towards a {{Sound Construction}} of {{EVM Bytecode Control-Flow Graphs}}},
  booktitle = {Proceedings of the 26th {{ACM International Workshop}} on {{Formal Techniques}} for {{Java-like Programs}} (FTfJP)},
  author    = {Arceri, Vincenzo and Merenda, Saverio Mattia and Dolcetti, Greta and Negrini, Luca and Olivieri, Luca and Zaffanella, Enea},
  year      = 2024,
  pages     = {11--16}
}

@article{ayubSoundAnalysisMigration2024,
  title   = {Sound Analysis and Migration of Data from {{Ethereum}} Smart Contracts},
  author  = {Ayub, Maha and Khan, Muhammad Waiz and Janjua, Muhammmad Umar},
  year    = 2024,
  journal = {Automated Software Engineering},
  volume  = {31},
  number  = {1},
  pages   = {21}
}

@inproceedings{caiEnablingCompleteAtomicity2024,
  title     = {Enabling {{Complete Atomicity}} for {{Cross-Chain Applications Through Layered State Commitments}}},
  booktitle = {Proceedings of the 43rd {{International Symposium}} on {{Reliable Distributed Systems}} ({{SRDS}})},
  author    = {Cai, Yuandi and Cheng, Ru and Zhou, Yifan and Zhang, Shijie and Xiao, Jiang and Jin, Hai},
  year      = 2024,
  pages     = {248--259}
}

@article{wang2024ContractsentryStaticAnalysis,
  title   = {Contractsentry: A Static Analysis Tool for Smart Contract Vulnerability Detection},
  author  = {Wang, Shiji and Zhao, Xiangfu},
  year    = 2024,
  journal = {Automated Software Engineering},
  volume  = {32},
  number  = {1},
  pages   = {1}
}

@inproceedings{sun2025FIRESmartContract,
  title     = {{{FIRE}}: {{Smart Contract Bytecode Function Identification}} via {{Graph-Refined Hybrid Feature Encoding}}},
  booktitle = {Proceedings of the 16th {{International Conference}} on {{Internetware}} (Internetware)},
  author    = {Sun, Yu and Bao, Lingfeng and Yang, Xiaohu},
  year      = 2025,
  pages     = {378--388}
}

@article{grossmanPracticalVerificationSmart2024,
  title   = {Practical {{Verification}} of {{Smart Contracts}} Using {{Memory Splitting}}},
  author  = {Grossman, Shelly and Toman, John and Bakst, Alexander and Arora, Sameer and Sagiv, Mooly and Nandi, Chandrakana},
  year    = 2024,
  journal = {Artifact for our paper titled "Practical Verification Of Smart Contracts using Memory Splitting"},
  volume  = {8},
  number  = {OOPSLA2},
  pages   = {356:2402--356:2433}
}

@article{hanOSwapPreservingAtomicity2026,
  title   = {{{OSwap}}: {{Preserving}} the {{Atomicity}} and {{Indistinguishability}} of \textbackslash bm N\_\textbackslash bm 1\textbackslash bm \textbackslash sim \textbackslash bm N\_\textbackslash bm 2n1{$\sim$}n2 {{Swap Without Synchronous Blockchain Communication}}},
  author  = {Han, Panpan and Yan, Zheng and Yang, Laurence T. and Bertino, Elisa},
  year    = 2026,
  journal = {IEEE Transactions on Dependable and Secure Computing},
  volume  = {23},
  number  = {1},
  pages   = {477--490}
}

@inproceedings{heCodeNotNatural2024,
  title     = {Code Is Not {{Natural Language}}: {{Unlock}} the {{Power}} of {{Semantics-Oriented Graph Representation}} for {{Binary Code Similarity Detection}}},
  booktitle = {Proceedings of the 33rd {{USENIX Security Symposium}} ({{USENIX Security}} 24)},
  author    = {He, Haojie and Lin, Xingwei and Weng, Ziang and Zhao, Ruijie and Gan, Shuitao and Chen, Libo and Ji, Yuede and Wang, Jiashui and Xue, Zhi},
  year      = 2024,
  pages     = {1759--1776}
}

@inproceedings{huangAdvancingWeb302024,
  title     = {Advancing {{Web}} 3.0: {{Making Smart Contracts Smarter}} on {{Blockchain}}},
  booktitle = {Proceedings of the {{ACM Web Conference}} 2024 (WWW)},
  author    = {Huang, Junqin and Kong, Linghe and Cheng, Guanjie and Xiang, Qiao and Chen, Guihai and Huang, Gang and Liu, Xue},
  year      = 2024,
  pages     = {1549--1560}
}

@article{jiaoSurveyEthereumSmart2024,
  title   = {A {{Survey}} of {{Ethereum Smart Contract Security}}: {{Attacks}} and {{Detection}}},
  author  = {Jiao, Tengyun and Xu, Zhiyu and Qi, Minfeng and Wen, Sheng and Xiang, Yang and Nan, Gary},
  year    = 2024,
  journal = {Distributed Ledger Technologies: Research and Practice},
  volume  = {3},
  number  = {3},
  pages   = {23:1--23:28}
}

@inproceedings{kumarVulnerabilitiesSmartContracts2024,
  title     = {``{{Vulnerabilities}} in {{Smart Contracts}}: {{A Detailed Survey}} of {{Detection}} and {{Mitigation Methodologies}}''},
  booktitle = {Proceedings of the 2024 {{International Conference}} on {{Emerging Technologies}} in {{Computer Science}} for {{Interdisciplinary Applications}} ({{ICETCS}})},
  author    = {Kumar, Nayantara K and Honnungar, Niranjan V and Sharwari Prakash, M and Lohith, J J},
  year      = 2024,
  pages     = {1--7}
}

@article{liASTRODetectingAccess2025,
  title   = {{{ASTRO}}: {{Detecting Access Control Vulnerabilities}} in {{Smart Contracts}} via {{Graph Similarity Comparison}}},
  author  = {Li, Wei and Nan, Yuhong and Ye, Mingxi and Zhang, Jingwen and Zheng, Peilin and Zheng, Zibin},
  year    = 2025,
  journal = {IEEE Transactions on Software Engineering},
  volume  = {51},
  number  = {12},
  pages   = {3267--3283}
}

@inproceedings{liDemoEnhancingSmart2024,
  title     = {Demo: {{Enhancing Smart Contract Security Comprehensively}} through {{Dynamic Symbolic Execution}}},
  booktitle = {Proceedings of the 2024 on {{ACM SIGSAC Conference}} on {{Computer}} and {{Communications Security}} (CCS)},
  author    = {Li, Zhaoxuan and Zhao, Ziming and Li, Wenhao and Zhang, Rui and Xue, Rui and Lu, Siqi and Zhang, Fan},
  year      = 2024,
  pages     = {5072--5074}
}

@inproceedings{priftiSmartContractVulnerability2024,
  title     = {Smart {{Contract Vulnerability Detection Using Deep Learning Algorithms}} on {{EVM}} Bytecode},
  booktitle = {Proceedings of the 13th {{Mediterranean Conference}} on {{Embedded Computing}} ({{MECO}})},
  author    = {Prifti, Lejdi and Cico, Betim and Karras, Dimitrios},
  year      = 2024,
  pages     = {1--7}
}

@article{suDiSCoDecompilingEVM2025,
  title   = {{{DiSCo}}: {{Towards Decompiling EVM Bytecode}} to {{Source Code}} Using {{Large Language Models}}},
  author  = {Su, Xing and Liang, Hanzhong and Wu, Hao and Niu, Ben and Xu, Fengyuan and Zhong, Sheng},
  year    = 2025,
  journal = {Proceedings of the ACM on Software Engineering},
  volume  = {2},
  number  = {FSE},
  pages   = {FSE103:2311--FSE103:2334}
}

@article{wangContractCheckCheckingEthereum2024,
  title   = {{{ContractCheck}}: {{Checking Ethereum Smart Contracts}} in {{Fine-Grained Level}}},
  author  = {Wang, Xite and Tian, Senping and Cui, Wei},
  year    = 2024,
  journal = {IEEE Transactions on Software Engineering},
  volume  = {50},
  number  = {7},
  pages   = {1789--1806}
}

@article{wangEfficientlyDetectingReentrancy2024,
  title   = {Efficiently {{Detecting Reentrancy Vulnerabilities}} in {{Complex Smart Contracts}}},
  author  = {Wang, Zexu and Chen, Jiachi and Wang, Yanlin and Zhang, Yu and Zhang, Weizhe and Zheng, Zibin},
  year    = 2024,
  journal = {Proceedings of the ACM on Software Engineering},
  volume  = {1},
  number  = {FSE},
  pages   = {8:161--8:181}
}

@article{wangEmpiricalAnalysisSmart2026,
  title   = {Empirical {{Analysis}} of {{Smart Contract Factories}} on {{EVM-compatible Chains}}},
  author  = {Wang, Ziyue and Shen, Zongwen and Chen, Lei and Song, Wei and Ge, Jidong and Huang, LiGuo and Luo, Bin},
  year    = 2026,
  journal = {ACM Transactions on Software Engineering and Methodology}
}

@article{weiSurveyQualityAssurance2024,
  title   = {Survey on {{Quality Assurance}} of {{Smart Contracts}}},
  author  = {Wei, Zhiyuan and Sun, Jing and Zhang, Zijian and Zhang, Xianhao and Yang, Xiaoxuan and Zhu, Liehuang},
  year    = 2024,
  journal = {ACM Computing Surveys},
  volume  = {57},
  number  = {2},
  pages   = {32:1--32:36}
}

@article{zhangEVMShieldInContractState2024,
  title   = {{{EVM-Shield}}: {{In-Contract State Access Control}} for {{Fast Vulnerability Detection}} and {{Prevention}}},
  author  = {Zhang, Xiaoli and Sun, Wenxiang and Xu, Zhicheng and Cheng, Hongbing and Cai, Chengjun and Cui, Helei and Li, Qi},
  year    = 2024,
  journal = {IEEE Transactions on Information Forensics and Security},
  volume  = {19},
  pages   = {2517--2532}
}

@inproceedings{zhangInferringLikelyCountingrelated2025,
  title     = {Inferring {{Likely Counting-related Atomicity Program Properties}} for {{Persistent Memory}}},
  booktitle = {Proceedings of the 2025 {{USENIX Annual Technical Conference}} ({{USENIX ATC}} 25)},
  author    = {Zhang, Yunmo and Qiu, Junqiao and Xu, Hong and Xue, Chun Jason},
  year      = 2025,
  pages     = {1441--1450}
}

@article{zhuSurveySecurityAnalysis2024,
  title   = {A {{Survey}} on {{Security Analysis Methods}} of {{Smart Contracts}}},
  author  = {Zhu, Huijuan and Yang, Lei and Wang, Liangmin and Sheng, Victor S.},
  year    = 2024,
  journal = {IEEE Transactions on Services Computing},
  volume  = {17},
  number  = {6},
  pages   = {4522--4539}
}

@inproceedings{boi2024VulnHuntGPTSmartContract,
  title     = {{{VulnHunt-GPT}}: A {{Smart Contract}} Vulnerabilities Detector Based on {{OpenAI chatGPT}}},
  booktitle = {Proceedings of the 39th {{ACM}}/{{SIGAPP Symposium}} on {{Applied Computing}}},
  author    = {Boi, Biagio and Esposito, Christian and Lee, Sokjoon},
  year      = 2024,
  pages     = {1517--1524}
}

@inproceedings{hu2023LargeLanguageModelPowered,
  title     = {Large {{Language Model-Powered Smart Contract Vulnerability Detection}}: {{New Perspectives}}},
  booktitle = {2023 5th {{IEEE International Conference}} on {{Trust}}, {{Privacy}} and {{Security}} in {{Intelligent Systems}} and {{Applications}} ({{TPS-ISA}})},
  author    = {Hu, Sihao and Huang, Tiansheng and {\.I}lhan, Fatih and Tekin, Selim Furkan and Liu, Ling},
  year      = 2023,
  pages     = {297--306}
}

@article{wei2025AdvancedSmartContract,
  title   = {Advanced {{Smart Contract Vulnerability Detection}} via {{LLM-Powered Multi-Agent Systems}}},
  author  = {Wei, Zhiyuan and Sun, Jing and Sun, Yuqiang and Liu, Ye and Wu, Daoyuan and Zhang, Zijian and Zhang, Xianhao and Li, Meng and Liu, Yang and Li, Chunmiao and Wan, Mingchao and Dong, Jin and Zhu, Liehuang},
  year    = 2025,
  journal = {IEEE Transactions on Software Engineering},
  volume  = {51},
  number  = {10},
  pages   = {2830--2846}
}

\end{document}